\documentclass[11pt,a4paper]{article}

\usepackage[truedimen,margin=35mm]{geometry} 

\usepackage{mathrsfs}
\usepackage{amssymb}
\usepackage{amsmath}
\usepackage{ascmac}
\usepackage{amsthm}
\usepackage[pdftex]{graphicx}
\usepackage[pdftex]{color}
\usepackage{natbib}
\usepackage{setspace}
\usepackage{color}
\usepackage{url}

\usepackage{titlesec}
\titleformat*{\section}{\large\bfseries}
\titleformat*{\subsection}{\it}

\def\ep{{\varepsilon}}
\def\la{{\lambda}}

\def\gam{\gamma}

\title{{\bf Bayesian Semiparametric Modeling of Response Mechanism for Nonignorable Missing Data}}
\date{}

\begin{document}

\maketitle
\doublespacing

\vspace{-1.5cm}
\begin{center}
{\large Shonosuke Sugasawa$^1$, Kosuke Morikawa$^2$ and Keisuke Takahata$^3$}
\end{center}

\medskip
\noindent
$^1$Center for Spatial Information Science, The University of Tokyo\\
$^2$Graduate School of Engineering Science, Osaka University\\
$^3$Graduate School of Economics, Keio University

\vspace{5mm}
\begin{center}
{\bf \large Abstract}
\end{center}
Statistical inference with nonresponse is quite challenging, especially when the response mechanism is nonignorable. In this case, the validity of statistical inference depends on untestable correct specification of the response model. To avoid the misspecification, we propose semiparametric Bayesian estimation in which an outcome model is parametric, but the response model is semiparametric in that we do not assume any parametric form for the nonresponse variable. We adopt penalized spline methods to estimate the unknown function. We also consider a fully nonparametric approach to modeling the response mechanism by using radial basis function methods. Using P\'{o}lya-gamma data augmentation, we developed an efficient posterior computation algorithm via Gibbs sampling in which most full conditional distributions can be obtained in familiar forms. The performance of the proposed method is demonstrated in simulation studies and an application to longitudinal data.


\newpage
Handling missing data in inappropriate ways may lead to crucial selection bias in data analysis. In particular, the specification of the response mechanism is essential for analyzing such data. If the response mechanism is misspecified, the statistical inference on the parameter of interest would be seriously biased. Nevertheless, it is often assumed that the response mechanism is ignorable or missing at random (MAR) because the assumption does not require any specification for the response mechanism \cite{Rubin1976}. Moreover, there have been several useful methods for ignorable missing data enjoying nice properties such as the double robustness \cite{Robins1994,Kang2007} and the multiple robustness \cite{Han2014}. In real data analysis, however, there are many unacceptable situations to believe the ignorability. Therefore, it is requisite to develop a method for analyzing nonignorable or missing not at random (MNAR) data \cite{Little2002}.

To analyze MNAR data, (i) a response model, which is a parametric model of the response mechanism,  needs to be correctly specified as well as (ii) the outcome model \cite{Greenlees1982,Diggle1994}. It has been criticized to analyze missing data under the MNAR assumption due to the strong assumption, so that several types of semiparametric models have been considered. \cite{Tang2003} and \cite{Zhao2015} proposed a semiparametric estimator for the outcome model without specifying any response model by using an instrumental variable. On the contrary, \cite{Qin2002,Chang2008}, and \cite{Kott2010} proposed a semiparametric estimator for the response model without specifying any outcome model. 
There are some literatures regarding Bayesian approaches to MNAR \cite{Durrant2006,Im2017} with parametric response models.

Recently, \cite{Kim2011} and \cite{Shao2016} proposed a semiparametric estimator for a semiparametric response model. In the semiparametric response model, terms on observed variables are modeled in a nonparametric way whereas the unobserved variable is still a simple linear function. However, we can not generally know or expect the effect of the unobserved variable. For example, if a survey of income is of our interest, and lower-income earners tend to refuse the item of income, the linear logistic model would be appropriate for the response model. However, if (super) higher-income earners also tend to refuse the item, quadratic or more complicated functions would be required. 
To address this issue, \cite{Sang2018} proposed a semiparametric response model with a nonparametric unobserved part.
They developed an EM algorithm to estimate unknown parameters in outcome models, but it requires numerical integration to profile out the nonparametric part in each iteration.
Thus, when we consider complicated outcome models such as random effects models considered in our example in Section \ref{sec:exm}, the profiling approach is not necessarily feasible.
Therefore, some alternative approaches would be needed in this context.

In this paper, we consider a semiparametric response model as considered in \cite{Sang2018} and adopt penalized spline methods to estimate the nonparametric part of the unobserved variables. 
We assign prior distributions for unknown parameters in the model and consider Bayesian inference by generating posterior samples, which enables us to obtain point estimates as well as measures of uncertainty such as credible intervals. 
Moreover, an important advantage of the Bayesian approach over frequentist approaches is that the proposed techniques can be applied to a variety of outcome models including random effects models adopted in Section \ref{sec:exm}.
This is because, in each iteration of the Markov Chain Monte Carlo algorithm, the unobserved response variables are augmented similarly to multiple imputation \cite{Rubin1978,Rubin1987}, so that the posterior of parameters in outcome models can be done as if all the response variables were observed.
To develop an efficient posterior computation algorithm, we employ a data augmentation technique known as P\'{o}lya-gamma augmentation \cite{Polson2013}, and derive Gibbs sampling in which most of the full conditional distributions are obtained in familiar forms. 
We also consider a fully nonparametric approach to the response model which would be a significant improvement over the existing models including those in \cite{Sang2018}.
In this case, we adopt radial basis function methods for the unknown part of auxiliary variables and derive a similar Gibbs sampling algorithm.

This paper is organized as follows.
In Section \ref{sec:method}, we provide details of the proposed semiparametric approach to the response model, including the posterior computation algorithm.
In Section \ref{sec:method2}, we introduce a fully nonparametric approach which is a slight extension of the methods in Section \ref{sec:method}.
In Sections \ref{sec:sim}, we carry out simulation studies to compare the performance of the proposed method with those of existing methods. 
In Section \ref{sec:exm}, we analyze longitudinal data from clinical trial regarding drug therapies for Schizophrenia.
Finally, some discussions are given in Section \ref{sec:dis}.
The R code is available at GitHub repository (\url{https://github.com/sshonosuke/MNAR-spline}).

\section{Semiparametric modeling for response mechanism}
\label{sec:method}

\subsection{Setup and semiparametric modeling}
Suppose that we are interested in estimating the parametric conditional distribution $f(y|x; \theta)$, where $y$ is a response, $x$ is a vector of covariates, and $\theta$ is a vector of unknown parameters.
For example, $f(y;x, \theta)$ can be normal density with mean $x^t\beta $ and variance $\sigma^2$, namely, $\theta=(\beta,\sigma^2)$ as a simple linear regression.
We assume that $x$ is always observed whereas $y$ is subject to missingness.
Let $s$ be the response indicator such that $s=1$ if $y$ is observed and $s=0$ otherwise.
We assume that $s$'s independently follow a Bernoulli distribution with the success probability $\pi(x,y)=P(s=1|x,y)$, which is referred to as the response mechanism.
In this article, we assume that the response mechanism is MNAR or nonignorable, that is, the response mechanism depends on the unobserved response.
Specifically, we consider the following response mechanism (response model):
\begin{equation}\label{selection}
P(s=1|x, y)=\psi(g^{\ast}(y)+z^t\delta),
\end{equation}
where $\psi(x)=\exp(x)/\{1+\exp(x)\}$ is the logistic function, $g^{\ast}(\cdot)$ is an unknown function and $z$ is a sub-vector of $x$. 
Define $x=(z^t, v^t)^t$, then $v$ is known as the nonresponse instrumental variable \cite{Wang2014}. 
The existence of the nonresponse instrumental variable guarantees the model identification of a semiparametric model defined in this subsection. The nonresponse instrumental variable is associated with the outcome conditional on the covariates but independent of the response indicator conditional on the covariates. Such a variable may be available in many empirical studies. For example, \cite{Miao2016} discussed the nonresponse instrumental variable in a study of the children's mental health in Connecticut \cite{Ibrahim2001,Zahner1992}.

Since $g^{\ast}(\cdot)$ is completely unspecified, the response model (\ref{selection}) is semiparametric.
For the estimation of $g^{\ast}(\cdot)$, we employ the P-spline of the form:
$$
g(y)=\phi_0+\sum_{j=1}^q\phi_jy^j+\sum_{\ell=1}^K\gamma_\ell(y-\kappa_{\ell})_+^q.
$$
Here $q$ is the degree of the spline, $(y-\kappa_{\ell})_+^q=(y-\kappa_{\ell})^qI_{\{y>\kappa_{\ell}\}}$, $\kappa_1<\ldots<\kappa_K$ is a set of fixed knots (whose choice will be discussed later) and $\phi=(\phi_0,\phi_1,\ldots,\phi_q)^t$ and $\gamma=(\gamma_1,\ldots,\gamma_K)^t$ are the coefficient vectors for the parametric part and the spline part, respectively.
If the knots are sufficiently spread over the range of $x$ and the number of knots $K$ is sufficiently large, then the class of functions $g(\cdot)$ can precisely approximate the unknown function $g^{\ast}(\cdot)$ even for small $q$, e.g. 2 or 3.
We here consider the case with fixed $K$ and locations of knots, but sensitivity analysis could be done in practice.
Since $q+K+1$ parameters are used in $g(y)$, we put a penalty on $\gamma$ by treating $\gamma$ as a random effect to avoid overfitting.
Specifically, we assume $\gamma\sim N(0,\la^{-1} I_K)$, where $\la$ is an unknown precision parameter to be estimated from the data.

\subsection{Posterior computation}\label{sec:pos}
We suppose the triplet $\{(x_i,y_i,s_i)\}$ is available for $i=1,\ldots,n$, where $n$ is the sample size.
The unknown parameters are $\theta$ in the outcome model, and $\phi,\gamma,\delta$ and $\lambda$ in the response model.
Let $\Xi$ be the collection of these unknown parameters.
The posterior distribution of $\Xi$ as well as missing observation $Y_{\rm mis}=\{y_{i} \mid s_{i}=0,i=1,\dots,n\}$ is given by
\begin{equation}\label{pos}
\begin{split}
\pi(\Xi,Y_{{\rm mis}}\ |\ {\rm Data}) & \propto\pi(\Xi)\prod_{i=1}^{n}\frac{\{\exp\{g(y_{i})+z_{i}^{t}\delta\}\}^{s_{i}}}{1+\exp\{g(y_{i})+z_{i}^{t}\delta\}}f(y_{i};x_{i},\theta),\\
 & =\pi(\Xi)\prod_{i=1}^{n}\frac{\{\exp(w_{1i}^{t}\phi+w_{2i}^{t}\gamma+z_{i}^{t}\delta)\}^{s_{i}}}{1+\exp(w_{1i}^{t}\phi+w_{2i}^{t}\gamma+z_{i}^{t}\delta)}f(y_{i};x_{i},\theta),
\end{split}
\end{equation}
where $w_{1i}=(1,y_i,\ldots,y_i^q)$, $w_{2i}=((y_i-\kappa_1)_+^q),\ldots,(y_i-\kappa_K)_+^q)$.
Using the P\'{o}lya-gamma data augmentation \cite{Polson2013}, we obtain the following augmented posterior:
\begin{equation}\label{pos2}
\begin{split}
&\pi(\Xi,Y_{\rm mis},\omega\ |\ {\rm Data})
\propto \pi(\Xi)\prod_{i=1}^n f(y_i;x_i,\theta)\exp\left\{\left(s_i-\frac12\right)u_i-\frac{\omega_iu_i^2}{2}\right\}p(\omega_i),
\end{split}
\end{equation} 
where $u_i\equiv u_i(y_i)=w_{1i}^t\phi+w_{2i}^t\gamma+z_i^t\delta$, and $p(\cdot)$ is a density function of the P\'{o}lya-gamma distribution ${\rm PG}(1,0)$.
Note that the integral with respect to $\omega_i$ reduces to the original posterior (\ref{pos}).
Under this expression, the conditional distribution of $u_i$ is normal.
As prior distributions on $\Xi$, we use multivariate normal distributions for $\phi$ and $\delta$, that is, $\phi\sim N(0,c_{\phi}^{-1}I_{q+1})$, $\delta\sim N(0,c_{\delta}^{-1}I_r)$, and a gamma distribution for $\lambda$, that is, $\lambda\sim {\rm Ga}(c_{\lambda},c_{\lambda})$, where ${\rm Ga}(a,b)$ denotes a gamma distribution with shape parameter $a$ and rate parameter $b$.
We adopt $c_{\phi}=c_{\delta}=10^{-4}$ and $c_{\lambda}=1$ as a default choice.  
Let $\pi(\theta)$ be a prior distribution for $\theta$, where its detailed form depends on the specific form of the outcome model $f(y;x,\theta)$.

To describe the sampling algorithm, we define $W_1=(w_{11},\ldots,w_{1n})^t$, $W_2=(w_{21},\ldots,w_{2n})^t$, $Z=(z_1,\ldots,z_n)^t$, $s_{\ast}=(s_1-1/2,\ldots,s_n-1/2)$, and $\Omega={\rm diag}(\omega_1,\ldots,\omega_n)$. 
The sampling algorithm is summarized as follows:

\begin{itemize}
\item[-]
(Sampling $\omega_i$) \ \ 
The full conditional distribution of $\omega_i$ is ${\rm PG}(1,u_i)$.
In general, a random variable having a P\'{o}lya-Gamma distribution ${\rm PG}(b,c)$ is expressed as 
$$
{\rm PG}(b,c)\stackrel{d}{=}\frac1{2\pi^2}\sum_{k=1}^{\infty}\frac{g_k}{(k-1/2)^2+c^2/(4\pi^2)}, \ \ \ \ g_k\sim {\rm Ga}(b,1),
$$  
where $g_k$'s are independent. 
Although a random sample of ${\rm PG}(b,c)$ can be generated by truncating the above infinite sum, \cite{Polson2013} developed an accept-reject algorithm, which is implemented in the R package \verb+pgdraw+ \cite{Makalic2016}.

\item[-]
(Sampling $\phi$) \ \ 
The full conditional density of $\phi$ is proportional to 
$$
\pi(\phi)\prod_{i=1}^n \exp\left\{\left(s_i-\frac12\right)w_{1i}^t\phi-\frac{\omega_i^2}{2}(w_{1i}^t\phi+w_{2i}^t\gamma+z_i^t\delta)^2\right\},
$$
thereby the full conditional distribution of $\phi$ is a multivariate normal distribution $N(A_{\phi}m_\phi,A_{\phi})$ with $A_\phi=(W_1^t\Omega W_1+c_{\phi}I_{q+1})^{-1}$ and $m_{\phi}=W_1^t\{s_{\ast}-\Omega(W_2\gamma+Z\delta)\}$.

\item[-]
(Sampling $\gamma$) \ \ 
Similarly to $\phi$, the full conditional distribution of $\gamma$ is a multivariate normal distribution $N(A_{\gam}m_\gam,A_{\gam})$, where $A_\gam=(W_2^t\Omega W_2+\lambda I_K)^{-1}$ and $m_{\phi}=W_2^t\{s_{\ast}-\Omega(W_1\phi+Z\delta)\}$.

\item[-]
(Sampling $\delta$) \ \ 
Similarly to $\phi$, the full conditional distribution of $\delta$ is a multivariate normal distribution $N(A_{\delta}m_\delta,A_{\delta})$, where $A_\delta=(Z^t\Omega Z+c_{\delta} I_r)^{-1}$ and $m_{\delta}=Z^t\{s_{\ast}-\Omega(W_1\phi+W_2\gam)\}$.

\item[-]
(Sampling $\lambda$) \ \
Generate $\lambda$ from its full conditional distribution given by ${\rm Ga}(c_\la+K,c_\la+\gamma^t\gamma)$.

\item[-]
(Sampling $y_i$ in $Y_{\rm mis}$) \ \ 
The full conditional distribution of $y_i$ is proportional to  
$$
g(y_i)\equiv f(y_i;x_i,\theta)\exp\left\{-\frac{u_i(y_i)}{2}-\frac{\omega_iu_i(y_i)^2}{2}\right\},
$$
which is an exponentially titled distribution, and is not a familiar form in general.
We adopt the Metropolis-adjusted Langevin Monte Carlo algorithm to update current values of $ y_i$.
With the current value $y_i$, the proposal $y_i^{\ast}$ is generated from the normal distribution $N(y_i-hV(y_i),2h)$, where $h$ is a user-specified step-size and  
\begin{align*}
&V(y_i)\equiv -\nabla \log g(y_i)\\
&=-\nabla \log f(y_i;x_i,\theta)+\left\{\frac12+\omega_iu_i(y_i)\right\}\bigg(\sum_{j=1}^qj\phi_jy_i^{j-1}+q\sum_{\ell=1}^K\gamma_\ell(y_i-\kappa_\ell)_+^{q-1}\bigg).
\end{align*}
Then, the proposal $y_i^{\ast}$ is accepted with probability
$$
\min\left\{1,\frac{g(y_i^{\ast})q_N(y_i;y_i^{\ast}-hV(y_i^{\ast}),2h)}{g(y_i)q_N(y_i^{\ast};y_i-hV(y_i),2h)}\right\},
$$
where $q_N(x;a,b)$ denotes the density function of $N(a,b)$.

\item[-]
(Sampling $\theta$) \ \ 
Since the augmented complete data is available, the full conditional posterior distribution of $\theta$ is proportional to $\pi(\theta)\prod_{i=1}^nf(y_i;x_i,\theta)$, which is the standard posterior distribution of $\theta$.
Hence, we could employ existing sampling techniques to update $\theta$ for the assumed outcome model $f(y_i;x_i,\theta)$.
\end{itemize}

Owing to the P\'{o}lya-gamma representation (\ref{pos2}), sampling steps for unknown parameters in the spline response model (\ref{selection}) are quite easy to carry out.
The full conditional distribution of $y_i$ in $Y_{\rm mis}$ is different from the assumed model $f(y_i;x_i,\theta)$ by the exponential term, which comes from the nonignorable response mechanism. 
When the response mechanism is MAR, the response model (\ref{selection}) is free from the unobserved value $y_i$ and $u_i$ does not depend on $y_i$, so that the full conditional distribution is the same as the assumed outcome model.

Finally, we address the way to select a suitable set of knots $\kappa_1<\cdots<\kappa_K$.
Provided that the knots are sufficiently spread out over the range of the response variable, the P-spline can approximate most smooth functions even under small $q$.
A crude way is to set $\kappa_1$ and $\kappa_K$ to low and high (e.g, $10\%$ and $90\%$) empirical quantiles of the observed responses, and set the other knots for equally spaced points between $\kappa_1$ and $\kappa_K$. 
However, such a strategy might fail under nonignorable missing since the missing value may take values out of the range of the observed responses and the response model depends on such missing values.
To address this issue, we consider two methods.
The first one is a similar adjustment to the crude method, that is, modify the crude vales of $\kappa_1$ and $\kappa_K$ to $\kappa_1^{\ast}=\kappa_1-a(\kappa_K-\kappa_1)/2$ and $\kappa_K^{\ast}=\kappa_K+a(\kappa_K-\kappa_1)/2$ for some positive constant $a$ specified by users.
As the second method, we consider a more data-adaptive way to deal with the idea, that is, we treat the positive constant $a$ as an unknown parameter and assign prior distributions to make the posterior inference. 
Since the knots appear in the posterior distributions (\ref{pos2}) through $u_i$, the full conditional posterior distribution of $a$ is proportional to 
$$
h(a)=\pi(a)\prod_{i=1}^n\exp\left\{\left(s_i-\frac{1}{2}\right)u_i-\frac{\omega_i u_i^2}{2}\right\},
$$
where $\pi(a)$ is the prior distribution on $a$.
To generate posterior samples from the non-familiar distribution, we simply adopt a random-walk MH algorithm which generates the proposal $a^{\ast}$ from a bivariate normal distribution $N(a^{\dagger},c)$ with current values $a^{\dagger}$ and a positive constant $c$, and accept the proposal with probability ${\rm min}\{1,h(a^{\ast})/h(a^{\dagger})\}$.
Note that in some applications (e.g. when larger response values are more likely to be missing), the use of the same $a$ for $\kappa_1^{\ast}$ and $\kappa_K^{\ast}$ would not be a reasonable strategy. 
Rather, we may set $\kappa_1^{\ast}=\kappa_1-a(\kappa_K-\kappa_1)/2$ and $\kappa_K^{\ast}=\kappa_K+b(\kappa_K-\kappa_1)/2$ with two positive constants, $a$ and $b$.
Under the setting, we can use similar strategies for generating posterior samples of $a$ and $b$ from their full conditional distributions.

\section{Fully nonparametric modeling for response mechanism}
\label{sec:method2}

The semiparametric response model (\ref{selection}) holds the parametric part of auxiliary variable $z$, which would be subject to misspecification. 
Although the effect of the misspecification seems limited under a situation where logistic response models can be seen from the relationship between distributions of observed and unobserved response variables \cite{Kim2011,Sang2018}, it would be useful to carry out more efficient statistical inference on parameters of interest.   
Thus, here we consider an extension of the semiparametric modeling in Section \ref{sec:method} to a fully nonparametric approach.
We consider the response model given by $P(s=1|y,x)=\psi(g(y)+h(z))$, where $h$ is the completely unknown function of $z$.
We adopt Gaussian radial basis function for estimating $h(z)$, that is, $h(z)$ is modeled by  
\begin{equation}\label{hz}
h(z)=\sum_{r=1}^R\xi_r\Phi(z; \eta_r),
\end{equation} 
where $\Phi(z; \eta_r)=\exp(-c_r\|z-\eta_r\|^2)$ with knot $\eta_r$ and scaling constant $c_r$, $\xi_r$ is an unknown coefficient, and $R$ is the number of radial basis functions. 
We note that the use of other radial basis functions does not change the following argument. 
Since full data of $z$ is available, the locations of knots $\eta_r$ can be readily determined by {\it k-means} algorithm with $R$ clusters. 
We assume  that $\xi_r\sim N(0, \lambda_{\xi}^{-1})$, independently for $r=1,\ldots,R$, to avoid over-fitting, where $\lambda_{\xi}$ is an unknown parameter playing a similar role to $\lambda$ in Section \ref{sec:method}.
We assign $\lambda_\xi\sim {\rm Ga}(c_{\xi}, c_{\xi})$ as a prior distribution with fixed hyperparameter $c_{\xi}$, where we set $c_{\xi}=1$ as the default choice.
Let $\xi=(\xi_1,\ldots,\xi_R)$ and $Z_{\Phi}$ be $(n,R)$-matrix with $(i,r)$-element given by $\Phi(z_i; \eta_r)$.

Under the nonparametric model with (\ref{hz}), the full conditional distribution of parameters and latent variables other than $\xi_r$ and $\lambda_{\xi}$ can be obtained by replacing $Z\delta$ by $Z_{\Phi}\xi$ in the algorithm given in Section \ref{sec:method}.
On the other hand, the full conditional distribution of $\xi$ is a multivariate normal distribution $N(A_{\xi}m_{\xi},A_{\xi})$, where $A_\xi=(Z_{\Phi}^t\Omega Z_{\Phi}+\lambda_{\xi} I_R)^{-1}$ and $m_{\delta}=Z_{\Phi}^t\{s_{\ast}-\Omega(W_1\phi+W_2\gam)\}$.
Also, the full conditional distribution of $\lambda_\xi$ is given by ${\rm Ga}(c_\xi+R,c_\xi+\xi^t\xi)$.

\section{Simulation study}\label{sec:sim}
We investigate the performance of the proposed method together with some existing methods.
To this end, we consider a simple linear regression model: 
\begin{equation}\label{DGP}
y_i=\beta_0+\beta_1x_{i1}+\beta_2x_{i2}+\ep_i, \ \ \ \ep_i\sim N(0,\sigma^2), \ \ \ \ i=1,\ldots,n,
\end{equation}
where $\beta\equiv (\beta_0,\beta_1,\beta_2)=(0.8,0.8,-0.5)$ and $\sigma^2=1$.
Here two covariates $x_{i1}$ and $x_{i2}$ were generated from a multivariate normal distribution, $(x_{i1},x_{i2})^t\sim N(0, \Sigma)$ with $(\Sigma)_{11}=(\Sigma)_{22}=1$ and $(\Sigma)_{12}=(\Sigma)_{21}=0.2$.
Based on the data generating model, we generated the response value $y_i$.
For the response mechanism, we independently generated the missing indicator $s_i$ from a Bernoulli distribution with the success probability $\pi_i$, and $y_i$ is observed/missing when $s_i=1$ or $0$.
In this study, we adopted 7 scenarios for $\pi_i$ given by 
\begin{align*}
&\text{(S1)} \ \  \pi_i={\rm logistic}(1.5-0.5y_i+0.2x_{i1}),\\
&\text{(S2)} \ \  \pi_i={\rm logistic}(2.5-0.2y_i-0.4y_i^2+0.2x_{i1}),\\
&\text{(S3)} \ \  \pi_i=1-\exp\{-\exp(2.5-0.2y_i-0.4y_i^2+0.2x_{i1})\},\\
&\text{(S4)} \ \  \pi_i={\rm logistic}(0.7y_i^2+0.2x_{i1}),\\
&\text{(S5)} \ \ \pi_i={\rm logistic}(0.5y_i^2+x_{i1}^2),\\
&\text{(S6)} \ \ \pi_i={\rm logistic}(1.5-2\sin y_i+0.2x_{i1}^2),\\
&\text{(S7)} \ \ \pi_i={\rm logistic}(0.7y_i^2+0.2x_{i2}).
\end{align*}
For each response mechanism, the overall response rates were ranging from $70\%$ to $80\%$. 
We considered two cases for the sample size, namely, $n=500$ and $n=1000$. Note that in this setup, the nonresponse instrumental variable is $x_{i2}$ except for Scenario 7 ($x_{1i}$ in Scenario 7), and in what follows, use $x_{2i}$ as the instrumental variable, i.e., response models are misspecified in Scenario 7.

We first focus on the population mean of the response variable, given by $\mu\equiv {\rm E}[{\rm E}[y_i|x_{i1},x_{i2}]]$.
For the estimation of $\mu$ based on the simulated missing data, the following methods are adopted. 

\begin{itemize}  
\item[-]
(OR: oracle method) \ 
It computes the average of the response values including nonresponse, thereby it cannot be adopted in practice.

\item[-]
(CC: complete-case method) \ 
As a crude method, we simply omit the nonresponse and calculate the sample average of observed responses. 

\item[-]
(LR: linear response method) \ 
We assume the logistic linear response model, $\pi_i={\rm logistic}(\phi_0+\phi_1y_i+\delta x_{i1})$, for the response mechanism and linear regression model (\ref{DGP}) for the outcome model to impute the missing outcome.
This is implemented in a Bayesian way.

\item[-] 
(SR: semiparametric response method)\ 
We apply the proposed semiparametric response model (\ref{selection}) combined with the linear regression (\ref{DGP}) as an outcome model. 
We set $z_i=x_{i1}$, $q=2$ (quadratic spline) and $K=10$.

\item[-]
(NR: nonparametric response method)\ 
We apply the proposed fully nonparametric response model given in Section \ref{sec:method2}, combined with the linear regression (\ref{DGP}) as an outcome model. 
We set $z_i=x_{i1}$, $q=2$ (quadratic spline), $K=10$, $R=K$ and $c_r=1$ (scaling constant in the radial basis).

\item[-]
(FI: fractional imputation method \cite{Riddles2016}) \  
The same logistic linear response model and outcome models as in LR are used.

\item[-]
(MM: weighted method-of-moments method \cite{Kott2010}) \ 
The same logistic linear response model and outcome models as in LR are used.

\item[-]
(SP: semiparametric profile likelihood method \cite{Sang2018}) \  
The assumed response model is a semiparametric function of $y_i$ and $x_{i1}$ as in the proposed SR method.
\end{itemize}

In the Bayesian methods, we generated 3000 posterior samples after discarding the first 2000 samples and computed the posterior mean of $n^{-1}\sum_{i=1}y_i$ as an estimator of $\mu$.
We evaluated the point estimates of $\mu$ using the square root of mean squared errors (RMSE) as well as bias based on 200 replications, where the results are given in Table \ref{tab:sim-mu}.

First, it shows that the CC method does not perform well when the missing is nonignorable. 
The performance of the parametric approaches, LR, FI, and MM, using a simple logistic linear model is quite reasonable in some scenarios, but it is not necessarily plausible as in Scenario 7. 
On the other hand, the proposed methods, SR and NR, perform well in almost all the scenarios although they can be inefficient under relatively simple missing structures such as Scenario 1.
Comparing the two methods, NR tends to be more inefficient than SR under situations where the parametric assumption for $x_{i1}$ in the response model is plausible as in Scenario 1$\sim $4 since NR adopted more complicated response models than SR.
However, under complicated missing structures as in Scenarios 5$\sim$7, NR tends to perform better than SR.
The frequentist version of the semiparametric approach, SP, shows plausible performance in Scenarios 6, but it does not necessarily work well in the other scenarios. 
Comparing the results with $n=500$ and $n=1000$, the difference of the performance of the OR method and the others tends to be smaller as $n$ increases possible because the larger number of $n$ enables us to precisely estimate the underlying missing structure.

\begin{figure}[!htb]
\centering
\includegraphics[width=13cm]{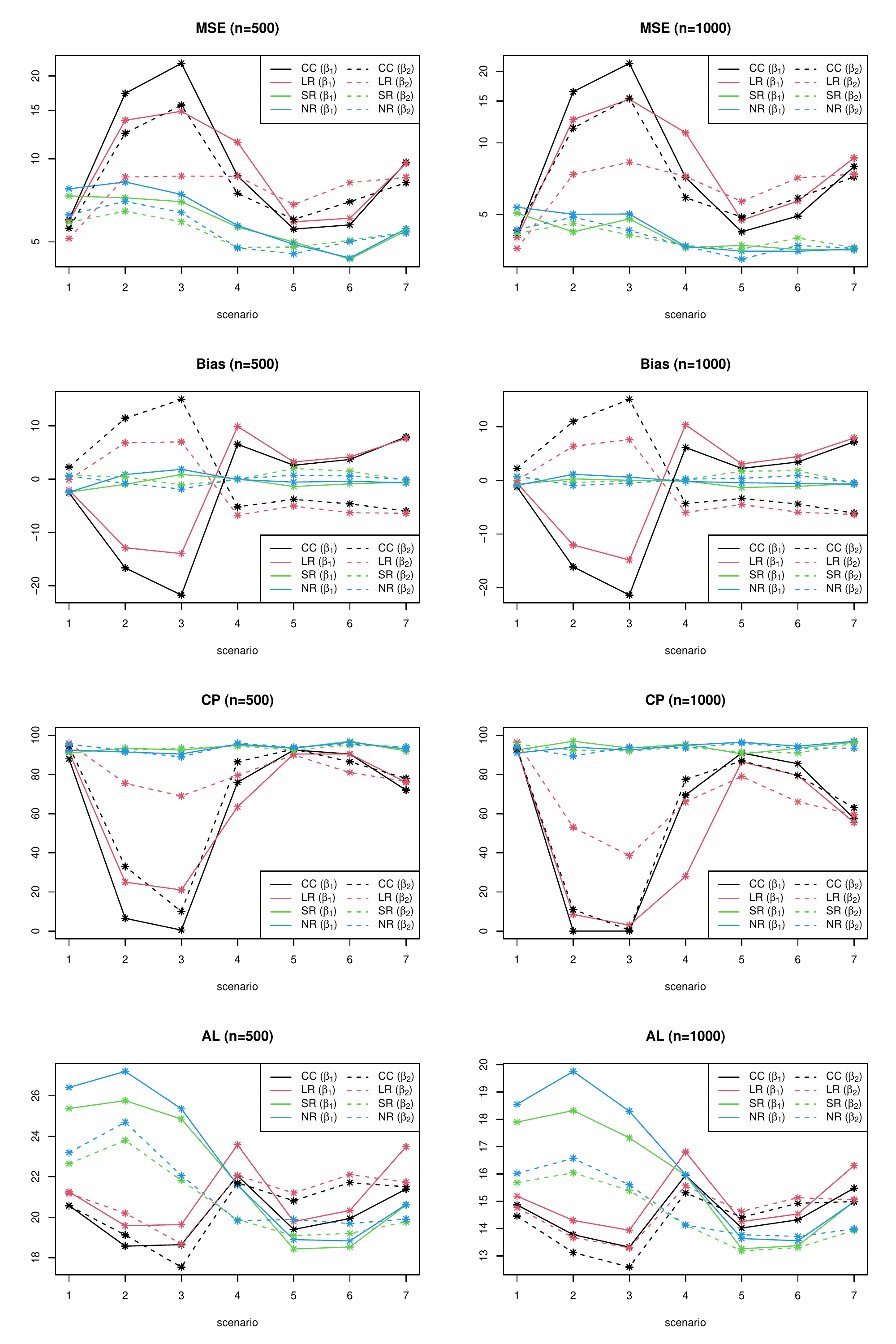}
\caption{Squared root of Mean squared errors (MSE) and bias of posterior means and coverage probabilities (CP) and average lengths (AL) of $95\%$ credible intervals of the regression coefficients based on complete-case (CC) method, logistic linear response (LR) model, the proposed semiparametric response (SR) and fully nonparametric response (NR) models under seven scenarios with $n=500$ and $n=1000$. 
All values are multiplied by 100.
\label{fig:sim-beta}
}
\end{figure}

We next investigated the performance in terms of estimating the regression parameters in the outcome models (\ref{DGP}).
We here mainly focus on the potential effects of the misspecification of the response model on estimating the regression parameters, thereby we adopted CC, LR, SR, and NR methods.
We computed point estimates and $95\%$ credible/confidence intervals of $\beta_1$ and $\beta_2$, respectively, and evaluated the performance by RMSE and bias of point estimates and coverage probabilities (CP) and average lengths (AL) of the intervals based on 200 replications.
We reported the results with $n=500$ and $n=1000$ in Figure \ref{fig:sim-beta}.

It is observed that LR shows preferable results in Scenario 1, but the performance in the other scenarios are not acceptable since the parametric linear assumption adopted in LR is misspecified in Scenarios $2\sim 7$.
It should be noted that CP can be quite low in some scenarios, which indicates that statistical inference under misspecification of the response model can break down. 
Regarding the proposed methods, they can be inefficient when the true response mechanism is simple as in Scenario 1, whereas they perform quite well in all the scenarios in terms of both point estimation (i.e. MSE and Bias) and posterior inference (i.e. CP and AL). 
Comparing SR and NR, NR is slightly more inefficient than SR when the parametric assumption concerning $x_{i1}$ is plausible as in Scenarios $1\sim 4$.
On the other hand, when the parametric assumption is violated as in Scenarios $5\sim 7$, NR tends to show better performance than SR. 
The amount of improvement of NR over SR in the case with $n=1000$ is larger than that of $n=500$ possibly because we can successfully estimate the complicated underlying response model under a large number of samples.

\begin{figure}[!htb]
\centering
\includegraphics[width=14cm]{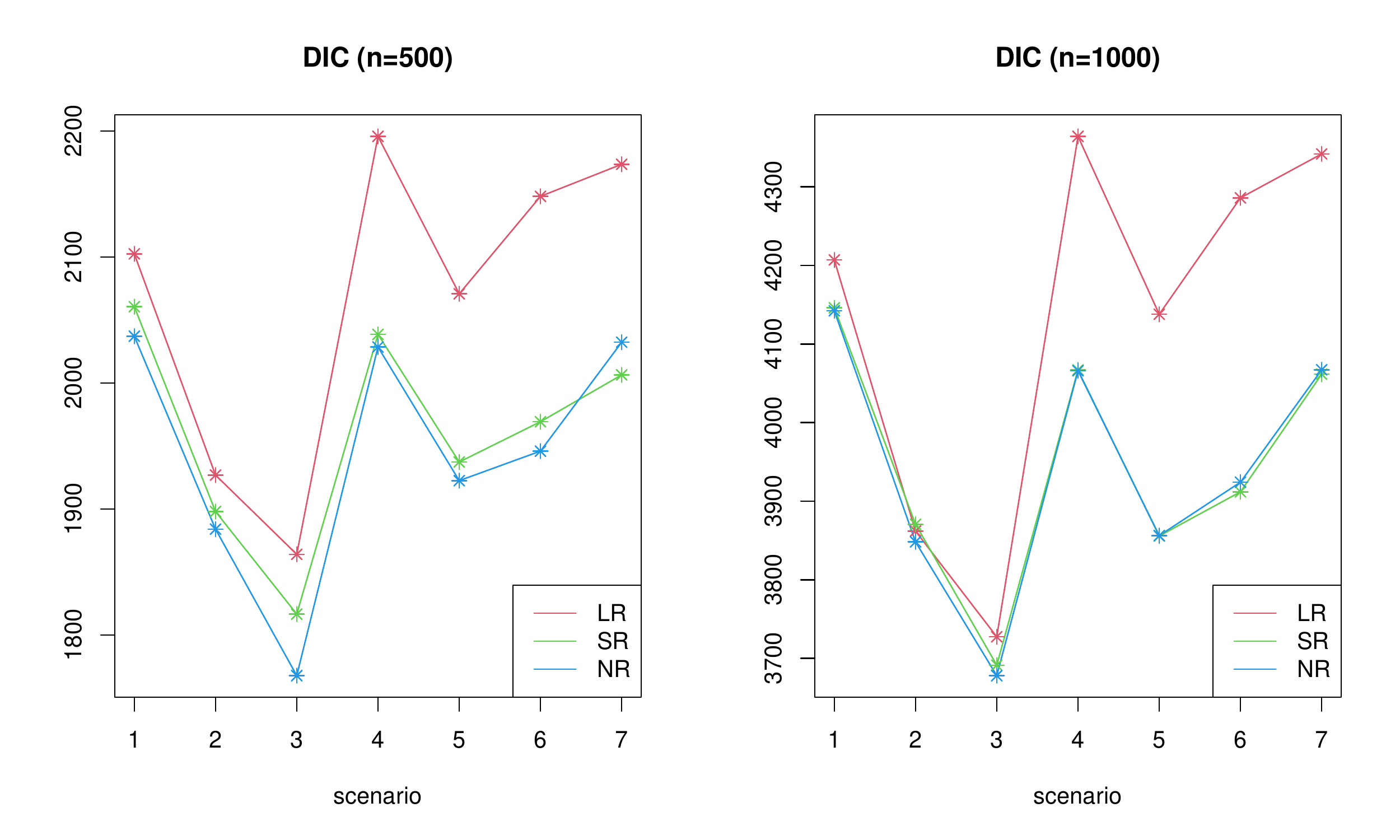}
\caption{Averaged values of deviance information criterion (DIC) of the logistic linear response (LR) model, the proposed semiparametric response (SR) and fully nonparametric response (NR) models under seven scenarios with $n=500$ and $n=1000$. 
\label{fig:sim-DIC}
}
\end{figure}

To assess relative goodness-of-fit, we computed deviance information criterion (DIC) \cite{spiegelhalter2002bayesian} of the three Bayesian methods (LR, SR, NR) based on the joint log-likelihood function 
$$
\sum_{i=1}^n\Big[\log \phi(y_i;x_i^t\beta, \sigma^2)+s_i\log \pi(y_i,x_i)+(1-s_i)\log\{1-\pi(y_i,x_i)\}\Big],
$$
where $x_i=(1,x_{i1},x_{i2})^t$ and $\pi(y_i,x_i)$ is a model of response probability.
In Figure \ref{fig:sim-DIC}, we present averaged values of the DIC under 7 scenarios. 
It is confirmed that DIC values are comparable when the simple LR model is not seriously misspecified (e.g. scenario $1\sim 3$).
On the other hand, DIC values of the proposed SR and NR methods are considerably smaller than those of the LR method when the underlying response mechanism cannot be approximated by the simple parametric model.

\begin{table}[htbp!]
\caption{Squared root of Mean squared errors (RMSE) and bias of point estimates of the population mean based on the oracle (OR) method, complete-case (CC) method, logistic linear response (LR) model, the proposed semiparametric response (SR) and fully nonparametric response (NR) models for response mechanism, fractional imputation (FI), weighted method-of-moments (MM) and semiparametric profile likelihood (SP) method in the seven scenarios with $n=500$ and $n=1000$.
All values are multiplied by 100. 
\label{tab:sim-mu}
}
\begin{center}
\begin{tabular}{cccccccccccccccccccc}
\hline
\multicolumn{9}{c}{ $n=500$ }\\
& Scenario   & OR & CC & LR & SR & NR & FI  & MM & SP \\
\hline
 & 1 & 4.71 & 18.49 & 8.37 & 10.81 & 13.44 & 9.33 & 9.08 & 10.95 \\
 & 2 & 4.23 & 27.93 & 6.34 & 7.44 & 9.31 & 12.88 & 8.08 & 19.24 \\
 & 3 & 4.43 & 36.25 & 6.53 & 7.07 & 7.87 & 23.59 & 9.96 & 25.44 \\
MSE & 4 & 4.06 & 23.33 & 7.46 & 5.82 & 6.07 & 7.52 & 7.53 & 8.96 \\
 & 5 & 6.84 & 18.06 & 8.45 & 8.72 & 7.96 & 10.31 & 8.76 & 11.49 \\
 & 6 & 6.54 & 10.51 & 8.44 & 8.74 & 8.30 & 8.38 & 9.03 & 6.94 \\
 & 7 & 7.00 & 25.70 & 18.83 & 12.17 & 12.00 & 18.13 & 18.02 & 16.23 \\
 \hline
 & 1 & 0.45 & -17.55 & 1.41 & 2.96 & 5.08 & 1.33 & 1.41 & -8.42 \\
 & 2 & 0.17 & -27.56 & 1.26 & 3.24 & 5.81 & -7.34 & 1.46 & -18.45 \\
 & 3 & 0.69 & -35.98 & -0.69 & 1.92 & 4.18 & -13.97 & 1.64 & -24.88 \\
Bias & 4 & 0.62 & 22.63 & -0.21 & 0.23 & -0.09 & 0.43 & 0.50 & 7.10 \\
 & 5 & 5.11 & 17.21 & 4.98 & 5.01 & 4.63 & 7.12 & 5.37 & 9.43 \\
 & 6 & 4.63 & -8.81 & 3.78 & 6.00 & 5.73 & 0.49 & 4.40 & 0.86 \\
 & 7 & 5.30 & 25.03 & 16.17 & 10.00 & 9.79 & 15.91 & 15.99 & 14.97 \\
 \hline
 \hline
 \multicolumn{9}{c}{ $n=1000$ }\\
& Scenario   & OR & CC & LR & SR & NR & FI  & MM & SP \\
\hline
 & 1 & 4.00 & 11.00 & 5.40 & 6.20 & 6.20 & 5.90 & 5.70 & 5.80 \\
 & 2 & 3.70 & 26.30 & 5.30 & 6.30 & 7.40 & 9.40 & 7.30 & 16.90 \\
 & 3 & 4.00 & 22.60 & 3.90 & 4.30 & 4.60 & 20.50 & 5.40 & 14.70 \\
MSE & 4 & 3.90 & 24.80 & 5.70 & 4.90 & 4.90 & 6.50 & 6.20 & 8.80 \\
 & 5 & 4.00 & 14.50 & 5.30 & 5.20 & 4.70 & 6.60 & 5.30 & 8.40 \\
 & 6 & 4.20 & 12.50 & 5.30 & 6.00 & 5.50 & 5.30 & 5.60 & 4.70 \\
 & 7 & 3.90 & 21.80 & 13.40 & 8.50 & 8.60 & 14.00 & 13.40 & 12.10 \\
 \hline
 & 1 & 2.40 & -10.30 & 2.50 & 3.00 & 3.40 & 2.70 & 2.60 & -4.00 \\
 & 2 & 2.00 & -26.10 & 2.20 & 3.50 & 4.90 & -3.30 & 3.00 & -16.30 \\
 & 3 & 2.30 & -22.40 & 1.40 & 2.20 & 2.90 & -19.10 & 1.90 & -14.30 \\
Bias & 4 & 2.20 & 24.30 & 1.00 & 1.20 & 0.90 & 1.60 & 1.80 & 7.50 \\
 & 5 & 2.10 & 13.90 & 2.20 & 1.90 & 2.00 & 4.10 & 2.20 & 6.70 \\
 & 6 & 2.60 & -11.80 & 2.30 & 4.10 & 3.60 & -0.60 & 2.70 & -0.70 \\
 & 7 & 2.20 & 21.40 & 11.90 & 6.90 & 6.90 & 12.60 & 12.20 & 11.10 \\
 \hline
\end{tabular}
\end{center}
\end{table}

\section{Example: Schizophrenia clinical trial}\label{sec:exm}
As a demonstration of the proposed method, we consider an application using a dataset of the randomized clinical trial of drug therapies for Schizophrenia, which is available from R package ``Surrogate".
In the trial, a placebo and treatment groups were compared, and the response of interest is an integer showing the severity of symptoms known as PANSS score, where high values indicate more severe symptoms. 
The patients were observed at weeks 1, 2, 4, 6 and 8 ($t = 1, 2, 3, 4, 5$) of the study.
In the dataset, 2151 patients are included and some patients have missing values.
If patients did not feel good enough to see doctors, they would not be able to see doctors and the corresponding values would be missing, thereby the response mechanism is considered as MNAR.
The overall missing rate is about $20\%$.
In this study, we are interested in the time-varying difference of PANSS scores between placebo and treatment groups.
Let $y_{it}$ denote the PANSS score for the $i$th individual at $t$th time, and we modeled the individual PANSS score as 
\begin{equation}\label{LMM}
y_{it}=\sum_{k\in \{0,1\}} R_{ik}(\beta_{0k}+\beta_{1k}T_t+\beta_{2k}T_t^2+\beta_{3k}T_t^3)+v_i+\ep_{ij},
\end{equation}
where $R_{ik}$ is the indicator whether the $i$th patient is included in Placebo ($k=0$) or Treatment ($k=1$) groups, $T_t$ denotes the measurement time, $v_i$ is an individual effect and $\ep_{ij}$ is an error term.
We assume that $v_i$ and $\ep_{ij}$ are mutually independent and distributed as $v_i\sim N(0,\tau^2)$ and $\ep_{ij}\sim N(0,\sigma^2)$. 
Note that in the model (\ref{LMM}), time change of the response variable $y_{it}$ is modeled by the measurement time separately for each group.
We let $s_{it}$ be the missing indicator such that $s_{it}=1$ if $y_{it}$ is observed, and $s_{it}=0$ otherwise. 
For the response mechanism, we employ the following model:
\begin{equation}\label{SM}
P(s_{it}=1|y_{it},s_{i,t-1})=\psi\Big(g(y_{it})+\delta_1 T_t+\delta_2 R_{i1}+\delta_3 s_{i,t-1}\Big),
\end{equation}
where $\psi(\cdot)$ is the logistic function and $s_{i0}=1$, and $g(\cdot)$ is a nonparametric function modeled by P-spline.
Note that the inclusion of $s_{i,t-1}$ in the response model addresses the time dependence of the missing indicator, and the joint distribution of $(s_{i1},\ldots,s_{iT})$ given the other variables is expressed as the product of the probability given in (\ref{SM}), so that we can still apply the same algorithm for posterior computation given in Section \ref{sec:method}.
Since $R_{i1}$ and $s_{i,t-1}$ are binary, and $T_t$ takes values on $\{1, 2, 4, 6, 8\}$, it would suffice to consider the semiparametric response (SR) model of the form (\ref{SM}) rather than the fully nonparametric model.

For the unknown parameters in the model (\ref{LMM}), we set priors described in Section \ref{sec:pos}, where the hyperparameters are specified as $c_{\beta}=c_{\phi}=c_{\delta}=10^{-4}$ and $c_{\sigma}=c_{\tau}=c_{\lambda}=1$.
As noted in Section \ref{sec:method}, the posterior sampling algorithm for unknown parameters in the outcome model is the same as one for complete data, thereby the posterior computation for the unknown parameters in (\ref{LMM}) can be easily implemented by using existing Gibbs sampling algorithm for linear mixed models \cite{Hobert1996}.

\begin{figure}[!htb]
\centering
\includegraphics[width=12cm]{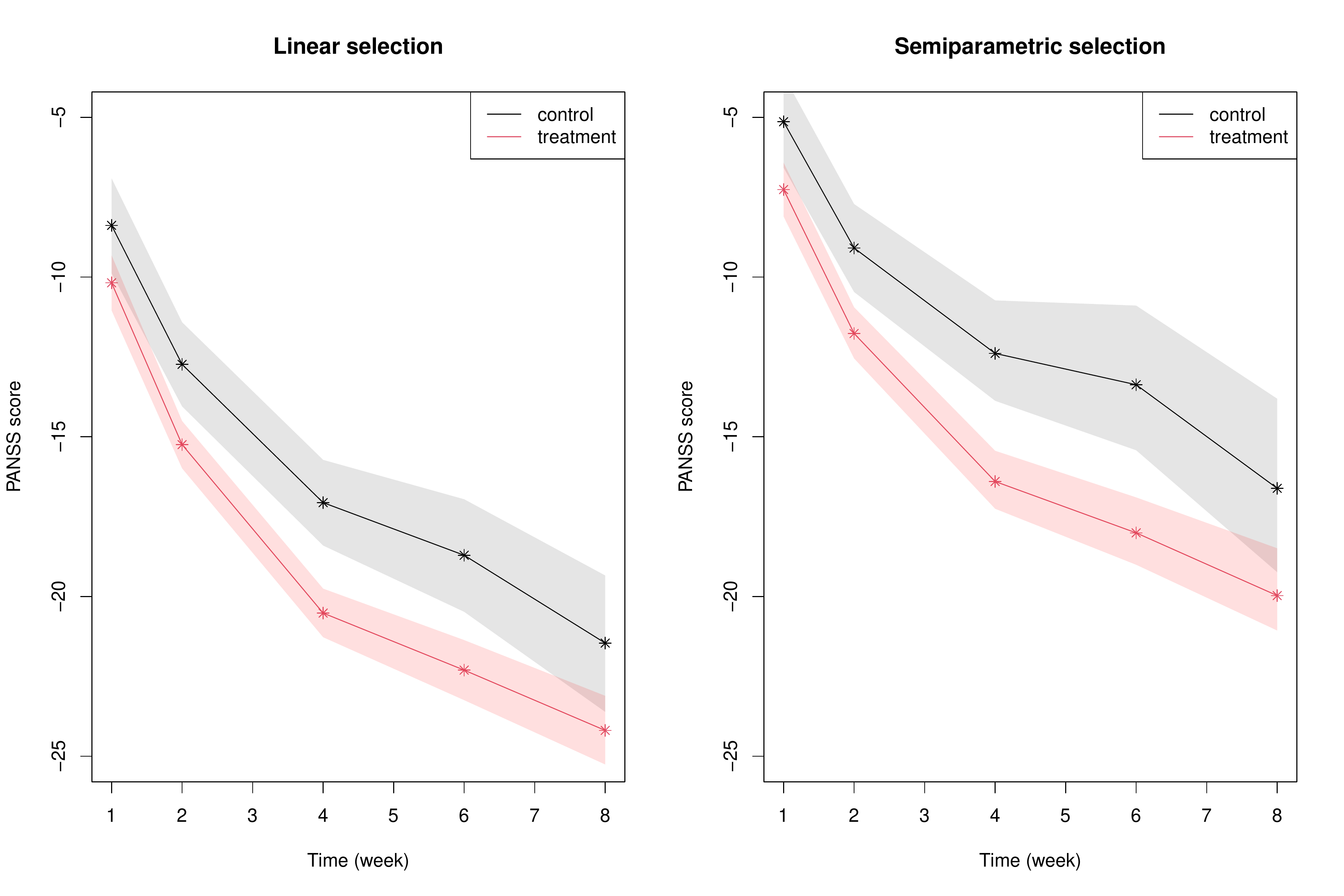}
\caption{Posterior means (solid lines) and point-wise $95\%$ credible intervals (dotted lines) of overall PANSS scores in two groups.}
\label{fig:PANSS}
\end{figure}

\begin{figure}[!htb]
\centering
\includegraphics[width=12cm]{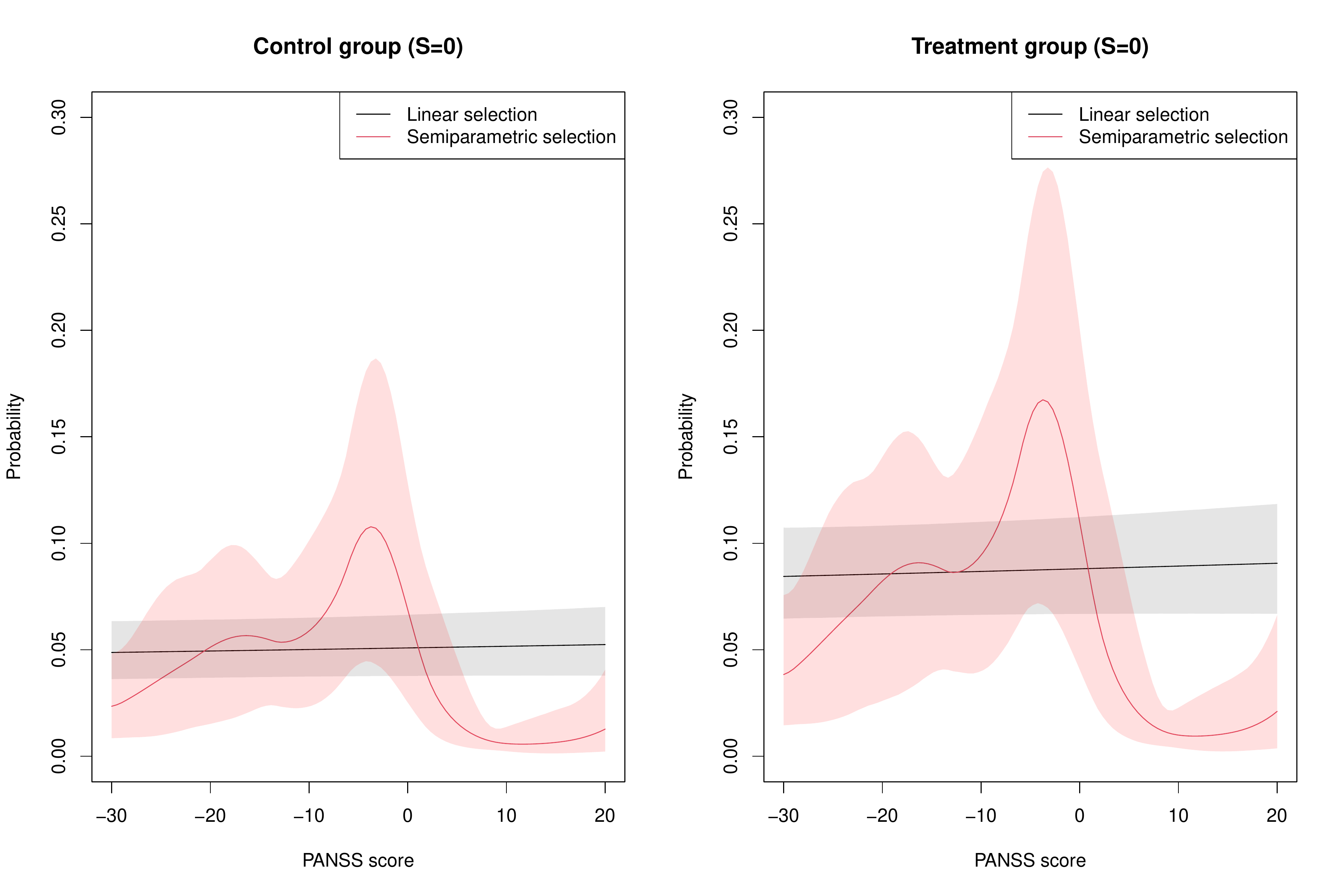}\\
\includegraphics[width=12cm]{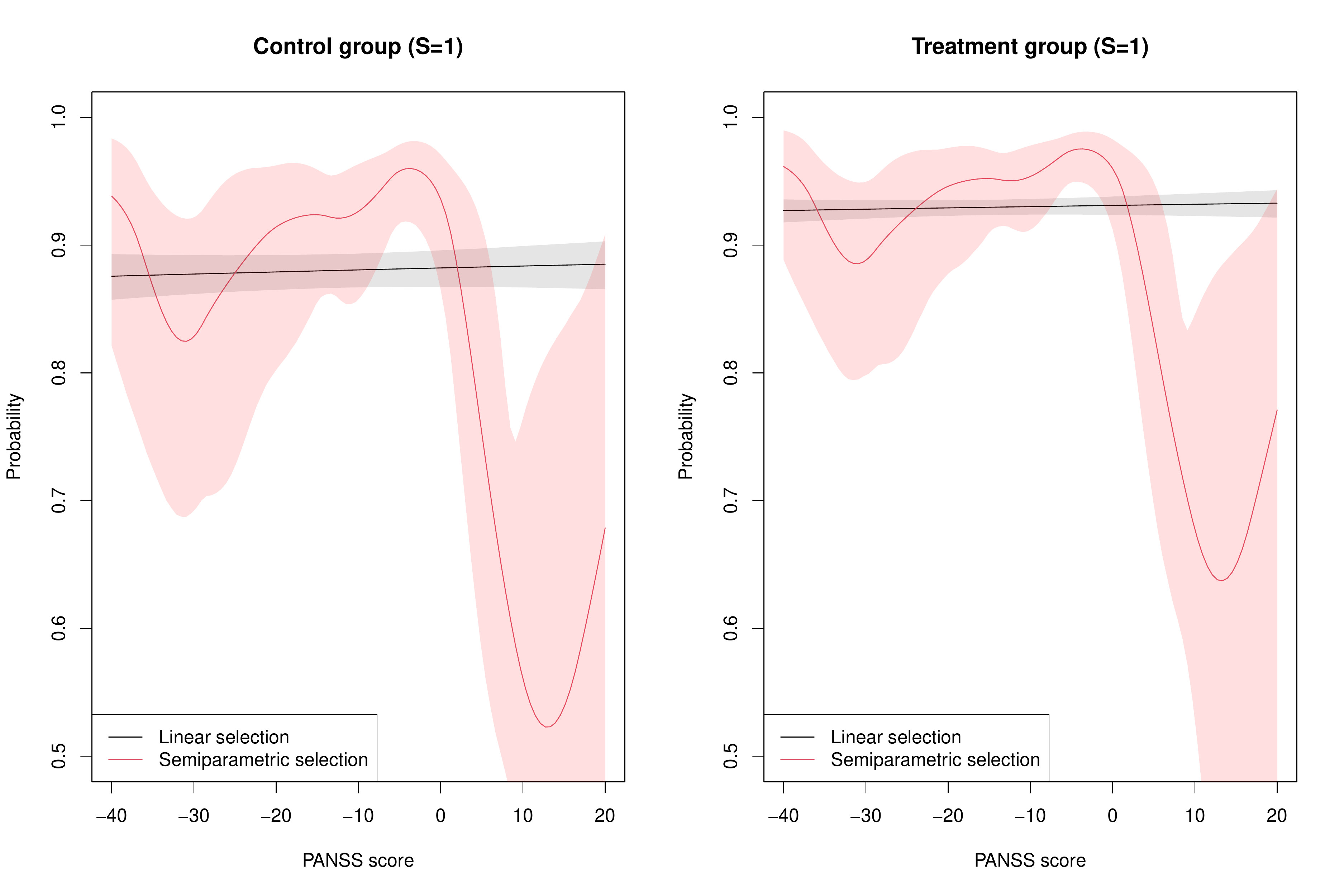}
\caption{Point-wise posterior means (solid lines) and $95\%$ credible intervals (dotted lines) of selection probabilities based on the standard linear response model and the proposed semiparametric for control and treatment group.  }
\label{fig:PANSS-mis}
\end{figure}

We considered the proposed method with $q=2$ and $K=10$.
For comparisons, we also applied the simple linear response (LR) model that replaces a linear function of $y_{it}$ with $g(\cdot)$ in (\ref{SM}).
We generated 40,000 posterior samples after discarding the first 10,000 samples as burn-in.
For model comparison, we computed DIC, and the value was 100,720 for LR and 97,566 for SR, which shows that SR is more suitable than LR for modeling the underlying response mechanism.
Based on the posterior samples, we computed the posterior means and point-wise $95\%$ credible intervals of regression lines in two (treatment and control) groups, which are reported in Figure \ref{fig:PANSS}.
It is observed that the estimates of regression lines are different between the proposed method and linear response model, and the credible intervals of the linear selection method are overlapped at some measurement times while those of the semiparametric selection model are slightly more separated at each measurement time.
Based on the results of the simulation study in Section \ref{sec:sim}, the result from the linear selection method is doubtful since it is subject to misspecification leading to serious bias in the estimation of parameters in outcome models.
We also computed the posterior means and point-wise $95\%$ credible intervals of the selection probability (\ref{SM}) as a function of PANSS score with $T_t=4$ and four combinations of $R_{i1}\in \{0,1\}$ and $s_{i,t-1}(=S)\in \{0,1\}$, which are presented in Figure \ref{fig:PANSS-mis}.
It is revealed that the effect of the treatment indicator on the missing probability is limited whereas the missing indicator of the previous time significantly changes the missing probability.
Also, the linear response model produces a very simple structure (almost constant over PANSS score) for missing probability while the proposed method seems to capture the underlying response mechanism flexibly as a complete function of PANSS score.
Such difference of flexibility in the estimation of missing probability would lead to the difference of resulting regression lines as shown in Figure \ref{fig:PANSS}.

Finally, we considered sensitivity check of the response model (\ref{SM}) and prior specifications.
To this end, we considered an alternative response model adding $T_t^2$ to (\ref{SM}), and two choices of $K\in \{10,20\}$.
We also considered alternative choice of the hyperparameter, $c_{\beta}=c_{\phi}=c_{\delta}=10^{-2}$ and $c_{\sigma}=c_{\tau}=c_{\lambda}=2$.
Based the same number of posterior samples, we computed the posterior means of the treatment effect at week $4$ and $8$, which are shown in Table \ref{tab:PANSS-sensitivity} and the results seem robust.

\begin{table}[htbp!]
\caption{Posterior means of the treatment effect at week 4 ($t=3$) and week $8$ (t=5) under various combinations of the tuning parameters and the response model.  
\label{tab:PANSS-sensitivity}
}
\begin{center}
\begin{tabular}{cccccccccccccccccccc}
\hline
$K$ & 10 & 15 & 10 & 15 & 10 & 10 & 10 \\
adding $T_t^2$ & False & False & True & True  &False  & False & False  \\
$c_{\beta}=c_{\phi}=c_{\delta}$ & $10^{-4}$ & $10^{-4}$ & $10^{-4}$ & $10^{-4}$ & $10^{-4}$ & $10^{-2}$ & $10^{-2}$ \\
$c_{\sigma}=c_{\tau}=c_{\lambda}$ & 1 & 1 & 1 & 1 & 2 & 2 & 1\\
week 4 & -4.01 & -4.01 & -3.99 & -4.00 & -3.93 & -4.17 & -4.31 \\
week 8 & -3.36 & -3.96 & -3.27 & -3.30 & -3.53 & -3.70 & -3.47 \\
\hline
\end{tabular}
\end{center}
\end{table}

\section{Discussion}\label{sec:dis}
This paper developed semiparametric Bayesian techniques under nonignorable missing responses, which can flexibly estimate the underlying response mechanism. 
We considered both semiparametric and nonparametric modeling for the response model and developed an efficient posterior computation algorithm using P\'{o}lya-gamma data augmentation. 
The advantage of the proposed Bayesian method is that it can be used with any outcome models since the posterior computation algorithm for parameters in the outcome model is the same as the case without missing responses.
We demonstrate the effectiveness of the proposed method through simulation studies and an application to longitudinal data.

For outcome models, we considered only a parametric model for simplicity, but we may employ semiparametric methods such as the generalized method of moments \cite{Yin2009}.
It should be remarked that the extension of the outcome modeling could be easily done since the posterior computation for outcome models is the same as that with complete data.
Since the detailed investigation would extend the scope of this paper, we left it to an interesting future study.

Finally, although we employed the logistic function as a link function in the response models due to its popularity in this context, we may use other link functions such as the probit link.
In such cases, the posterior computation algorithm given in Section \ref{sec:method} should be changed accordingly.

\section*{Acknowledgement}
This work is supported by Japan Society for the Promotion of Science (KAKENHI) grant numbers 18K12757 and 19K14592.

\bibliographystyle{chicago}
\bibliography{refs}

\end{document}